\documentclass[prb,twocolumn,floatfix,showpacs,superscriptaddress]{revtex4-1}
\usepackage{epsfig,subfigure,amssymb,amscd,amsmath,graphicx,amsfonts,mathtools,mathcomp,pbox,amssymb,bbm}
\usepackage{float}
\usepackage{silence}
\usepackage{times,longtable,math dots,hyperref}

\newcommand{\non}{\nonumber}

\newcommand{\bal}{\begin{align}}
\newcommand{\eal}{\end{align}}
\newcommand{\braket}[2]{\langle #1|#2\rangle}
\newcommand{\ket}[1]{     |    \,    #1    \rangle}
\newcommand{\bra}[1]{  \langle #1  \,  |}

\newcommand{\be}{\begin{equation}}
\newcommand{\ee}{\end{equation}}
\usepackage{tabularx}
\usepackage{todonotes}

\makeatletter
\newcommand*\dashline{\rotatebox[origin=c]{90}{$\dabar@\dabar@\dabar@$}}
\makeatother

\begin{document}

\title{Error generation and propagation in Majorana-based topological qubits}

\author{A. Conlon}
\email[For further information contact \! ]{Aaron.Conlon@mu.ie}
\affiliation{Department of Theoretical Physics, Maynooth University, Ireland.}
\author{D. Pellegrino}
\affiliation{Department of Theoretical Physics, Maynooth University, Ireland.}
\author{J. K. Slingerland}
\affiliation{Department of Theoretical Physics, Maynooth University, Ireland.}
\affiliation{Dublin Institute for Advanced  Studies, School of Theoretical  Physics, 10 Burlington Rd, Dublin, Ireland.}
\author{S. Dooley}
\affiliation{Dublin Institute for Advanced  Studies, School of Theoretical  Physics, 10 Burlington Rd, Dublin, Ireland.}
\author{G. Kells}
\affiliation{Dublin Institute for Advanced  Studies, School of Theoretical  Physics, 10 Burlington Rd, Dublin, Ireland.}

\begin{abstract}
We investigate dynamical evolution of a  topological memory that consists of two $p$-wave superconducting wires separated by a non-topological junction, focusing on the primary errors (i.e., qubit-loss) and secondary errors (bit and phase-flip) that arise due to non-adiabaticity. On the question of qubit-loss we examine the system's response to both periodic boundary driving and deliberate shuttling of the Majorana bound states. In the former scenario we show how the frequency dependent rate of qubit-loss is strongly correlated with the local density of states at the edge of wire, a fact that can make systems with a larger gap more susceptible to high frequency noise.  In the second scenario we  confirm previous predictions concerning super-adiabaticity and critical velocity, but see no evidence that the coordinated movement of edge boundaries reduces qubit-loss. Our analysis on secondary bit flip errors shows that it is necessary that non-adiabaticity occurs in both wires and that inter-wire tunnelling be present for this error channel to be open. We also demonstrate how such processes can be minimised by disordering central regions of both wires. Finally we identify an error channel for phase flip errors, which can occur due to mismatches in the energies of states with bulk excitations. In the non-interacting system considered here this error systematically opposes the expected phase rotation due to finite size splitting in the qubit subspace. 
\end{abstract}


\date{\today} \maketitle

\section{Introduction}
Quantum devices based on topological order are expected to be robust against many common forms of decoherence, making them an exciting prospect for quantum computation \cite{Kitaev2001,Kitaev2003,Kitaev2006,Nayak2008,Stanescu2016}. Although originally envisaged as a 2D platform, the realisation that quasi-1D proximity-coupled systems \cite{Fu2008,Lutchyn2010,Oreg2010}
could be employed in network architecture \cite{Sau2010,Alicea2011}  has led to a great deal of theoretical \cite{Duckheim2011,Chung2011,Choy2011,Kjaergaard2012,Martin2012,NadjPerge2013,Aasen2016} and experimental \cite{Mourik2012,Deng2012,Das2012,Finck2013,Churchill2013,Albrecht2016, Zhang2016,Deng2016,NadjPerge2014,Ruby2015,Pawlak2016,Zhang2018} work. However, despite significant advances, the current reality is that devices based on proximity-coupled superconductors still suffer from errors. There is a range of problematic processes that potentially limit the effectiveness of topological quantum memories. The majority of these\cite{Goldstein2011,Budich2012,Rainis2012,Mazza2013, Ho2014, Ng2015,Campbell2015,Pedrocchi2015,Huang2018, Knapp2018, KnappNoise} concern the inevitable interaction with an external environment. However research in this area has also focused on errors due to changes of the internal system parameters over a finite time, e.g., due to fluctuations in the underlying gate potentials \cite{Schmidt2012}, thermal noise\cite{Cheng2012}, or the deliberate shuttling of Majorana-Bound-States (MBS) around a network \cite{Cheng2011,Perfetto2013,Bauer2018,Karzig2013,Scheurer2013,Knapp2016,Rahmini2017,Karzig2015,KarzigShortcut}.

In this work we set out to examine error generation in this latter scenario and analyse how initial qubit-loss error can give rise to additional bit and phase-flip errors.  For idealised topological qubits, any excitations away from the ground state manifold represent a source of qubit-loss error. We examine this issue from the perspective of a simple topological memory constructed from two spinless $p$-wave superconducting wires, connected via a single tunnel junction, specifically studying how qubit-loss depends on (1) time-dependent fluctuations in the confining potentials and (2) the non-adiabatic transport of Majorana bound states. 

Our analysis in the case of oscillating boundary walls indicates that the frequency dependence of the error rate is strongly correlated with the local density of states at the edge of the wire.  We also find that the maximum coupling to bulk states is inversely proportional to the bulk gap, but that the frequency at which this maximum coupling occurs increases as we increase the superconducting gap. For the very low frequency oscillations this naturally coincides with the expected behaviour in the adiabatic limit. However on the high-frequency side we see that this can mean a reduction in qubit stability as the bulk gap is increased.

On the question of Majorana bound state transport, significant progress has been made previously by Refs. \onlinecite{Karzig2013,Scheurer2013} where it was shown that MBS can be transported without incurring any significant qubit loss, provided that the acceleration is sufficiently slow and the maximum velocity is less than a critical value. Our findings here are mostly in agreement with these previous works. However, we observe no advantage in keeping the wire's length fixed during a movement protocol, a  departure from previous predictions in relation to correlated movement of distant boundary walls outlined in Ref. \onlinecite{Scheurer2013}.
Our results also show that it is possible to think of  the instantaneous energy increase as a kinetic energy of the boundary wall, with an effective-mass $ M^* \propto k_F / 2 \Delta= k_F^2/2 E_{\text{gap}}$, that incurs relativistic-like corrections when it approaches the critical velocity. We also show that there is only a modest additional qubit-loss penalty for approaching the infinite acceleration limit in the $v_\text{max} < v_\text{crit}$ regime\footnote{This feature has been proposed to be used in so called bang-bang control protocols \cite{Karzig2015}.}.  

In addition we study how initial qubit-loss leads to secondary error processes such as bit and phase-flip error.  This is important because although qubit-loss the is the primary source of error in this setup, it is in principle detectable (e.g. via a projective energy measurement), and therefore its existence alone does not pose a fundamental problem for the stability of the quantum memory and/or any ongoing quantum computation. In contrast a more damaging situation occurs when the original error is followed by some secondary processes that, after some time, returns the system to the qubit-space with an error. These error processes cannot be detected without some additional error detection/correction protocols, a scenario which topological schemes are supposed to avoid.

\begin{figure}[t]
\centering
 	\includegraphics[width=1\columnwidth]{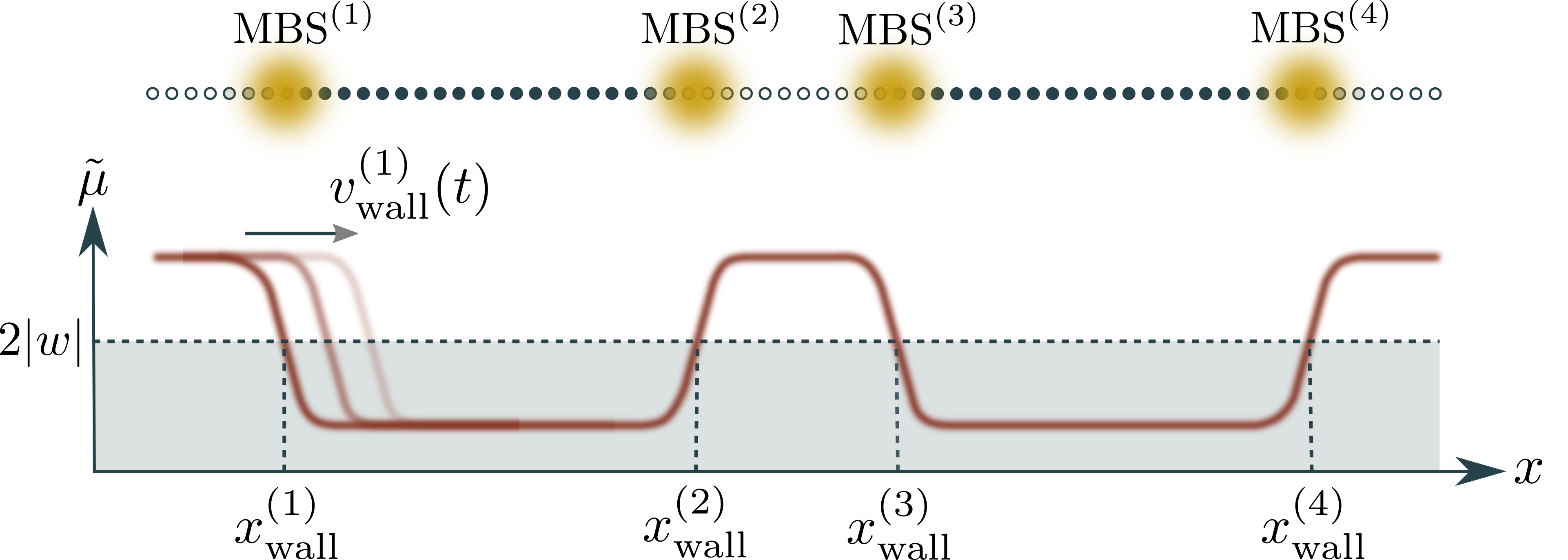}
	\caption{(Colour online) Top: A schematic illustration of the Kitaev chain used to encode a single topological qubit. Majorana bound states (MBS) are localised at the boundaries between topological and non-topological regions. Bottom: The chain is in a topological phase in regions where $|\tilde{\mu}| < 2|w|$ (the shaded area) and in a non-topological phase in regions where $|\tilde{\mu}| > 2|w|$. The MBS at $x_\text{wall}^{(i)}$ is manipulated by moving the boundary wall with a velocity profile $v_\text{wall}^{(i)}(t) = dx_\text{wall}^{(i)}(t)/dt$.}
        \label{fig::setup}
\end{figure}
For bit-flip error our central results are that boundary wall movement will generate localized quasi-particles that can propagate across the wire and cause errors if they tunnel through boundary walls separating topological regions. The times at which the errors begin to appear can be estimated from the velocity of the wave-front, which is approximately equal to the Fermi velocity  $v_f$. We also investigate a possible way to reduce such errors: by inducing disorder in the bulk of the wire, the bulk excitations can be made subject to disorder induced localization, which prevents propagation of the wave without lowering the critical velocity of the system.   

We also discuss undetectable phase errors. These could potentially occur in the same manner as the bit flip error, but we actually identify another process through which the qubit may be vulnerable to uncontrolled rotations and which can already occur in a single wire. The error process does not require any non-local effects and/or tunneling between topological domains, but it is dependent on the existence of an energy mismatch between certain bulk eigenstates. In our study, since we restrict ourselves only to non-interacting systems, this error channel is only open when there is some overlap between the edge modes, meaning that they are not strictly zero modes -- a typical finite size effect. This means in particular that we will already have some Rabi oscillations between the qubit states as they are not perfectly degenerate. These expected oscillations can be modified when the splitting of pairs of excited states differs from the qubit states' splitting. A similar error channel may be relevant also when one wishes to deliberately rotate non-topologically within the ground state manifold. In interacting systems, we might expect this type of phase error even if the system size is large enough to make the ground state splitting negligible, since the spectrum is no longer described by single particle modes and energy mismatches between excited states can occur even if the ground state degeneracy is robust. 

An outline of the paper is as follows. In section \ref{sect:Overview}, we review the Kitaev model for a $p$-wave superconducting wire and describe how we calculate and analyze the dynamics generated by the movement of boundary walls. In section \ref{sect:QubitLoss}, we discuss the qubit-loss error due to the moving walls. We divide this section into two parts. First, in section \ref{sect:oscillating_Qubit_Loss}, we focus on oscillating boundary walls. Next, in section \ref{sect:Adiabatic_Qubit_Loss} we investigate qubit-loss error due to non-adiabaticity during transport of Majorana bound states (MBS) along a wire. In section \ref{sect:secodaryerrors} we present our results on the secondary processes that can cause bit and phase-flip errors in the computational subspace. We also divide this in two: in section \ref{sect:BitflipError} we discuss bit-flip errors in the two-wire scenarios and in section \ref{sect:phase_error} we discuss phase errors due to non-adiabaticity. Finally, in section \ref{sect:Conclusions} we summarize and discuss possible avenues of future work. 

\section{Model}\label{sect:Overview}

\subsection{The Kitaev model}

The Hamiltonian for the Kitaev model of a $p$-wave superconductor is 
\begin{align}
\label{eq:Hamiltonian}
	H & = -\sum_{j=1}^{N} \tilde{\mu}_{j} (c^\dagger_{j }c_{j }-1/2) - w \sum_{j=1}^{N-1} (c^\dagger_{j}c_{j+1} + h.c.) \non \\ 
	& \quad + \tilde{\Delta} \sum_{j=1}^{N-1} ( c^\dagger_{j} c^\dagger_{j+1} + h.c. ) - \sum_{j=1}^{N} \lambda_{j}{c}^{\dag}_{j}c_{j},
\end{align}
where $c_j$ is the fermionic annihilation operator at the $j$'th lattice site, $\tilde{\mu}_{j}$ is the lattice chemical potential, $w$ represents the hopping strength between neighboring lattice sites, and $\tilde{\Delta} \neq 0$ is the lattice pairing parameter, which induces a superconducting gap. 

The chemical potential $\tilde{\mu}_j$ can vary across the wire and will be used to control the boundaries between topological regions (where $|\tilde{\mu}_j| < 2|w|$) and non-topological regions (where $|\tilde{\mu}_j| > 2|w|$). Majorana bound states are localised at the boundaries between topological and non-topological regions in the wire. We model a boundary wall separating a topological and a non-topological region -- at a position $x_\text{wall}$ along the wire -- by a sigmoid function for the chemical potential (see Fig.~\ref{fig::setup} for an illustration showing several boundary walls and see Appendix \ref{app::wall} for further details on the boundary wall function). To move a boundary wall we allow the wall position $x_\text{wall}(t)$ to be time-dependent. Equivalently, we can specify a time-dependent wall velocity $v_\text{wall}(t) = dx_\text{wall}(t)/dt$. This results in a time-dependent chemical potential $\tilde{\mu}_j(t)$, and hence a time-dependent Hamiltonian (\ref{eq:Hamiltonian}). Disorder in the chemical potential is introduced using the Gaussian random variable $\lambda_{j}$, which has zero mean $ \langle \lambda_{j} \rangle=0$ and variance $\langle \lambda_{i}\lambda_{j} \rangle = \alpha \delta_{ij}/a$, where $a$ is the lattice spacing. 

In order to solve the system, the Hamiltonian (\ref{eq:Hamiltonian}) can be rewritten as $H(t) = \sum_{n=1}^N \epsilon_n(t) [\beta_n^\dagger (t) \beta_n (t) - 1/2]$ via the Bogoliubov transformations:
\begin{equation}
\label{eq::bogoliubov}
\begin{split}
	& \beta_{n}(t) = \sum_{j} U_{j,n}(t) c_{j} + V^*_{j,n}(t) c^{\dagger}_{j},  \\
	& \beta^{\dagger}_{n}(t) = \sum_{j} U_{j,n}^{*}(t) c^{\dagger}_{j} + V_{j,n}(t) c_{j},
\end{split}
\end{equation}
where $U_{j,n}(t)$, $V_{j,n}(t)$ are the coefficients of the unitary matrix that diagonalizes the Bogoliubov-De Gennes (BdG) Hamiltonian associated with $H(t)$. We assume that the spectrum is labeled in increasing order $\epsilon_0 \leq ... \leq \epsilon_{N-1}$. Pairs of Majorana bound states, if well separated, are associated with zero energy modes ($\epsilon_0 \approx 0$, $\epsilon_1 \approx 0$,... etc.). If there are $m$ such zero modes $\{\beta_j\}_{j=0}^m$, then there are $2^m$ degenerate ground states $(\beta_0^\dagger)^{l_0}...(\beta_m^\dagger)^{l_m}|\text{GS}\rangle$, where $l_j \in \{0,1\}$ and $|\text{GS}\rangle$ is the BCS ground state.

We also note that the continuum limit of the chain can be obtained by taking the total number of lattice sites $N \rightarrow \infty$ and the lattice spacing $a \rightarrow 0$, with the length of the wire $L = Na$ kept constant. In this limit, the lattice Hamiltonian (\ref{eq:Hamiltonian}) becomes a low energy effective Hamiltonian 
\mbox{$H = \int_0^L \! \! dx\, \Phi^\dagger(x) [(-\partial_x^2/2m \!-\! \mu(x) \!-\! \lambda(x))\sigma_z \!+\! \Delta\partial_x \sigma_x] \Phi(x)$}, 
where 
\mbox{$\Phi^\dagger (x) = (c^\dagger (x) \: c(x))$}. 
It will be useful to express some of our results in terms of the continuum parameters $m$, $\mu$ and $\Delta$, which are related to the lattice parameters by $m = 1/(2wa^2)$, $\mu = \tilde{\mu}_{j} + 2w$, and $\Delta = 2a\tilde{\Delta}$. In this limit, the chain is in a topological phase when $\mu > 0$ and in a non-topological phase for $\mu < 0$. Using the continuum parameters, it is also useful to define the Fermi momentum $k_F = \sqrt{2m\mu}$, the Fermi velocity $v_F = k_{F}/m$, the gap energy $E_\text{gap} = \Delta k_F$, and the finite localization length due to disorder $l=v_F^2/\alpha$.

\subsection{The Topological qubit}\label{sec:topo_qubit}

Throughout this paper we consider a topological system that encodes the smallest unit of quantum information, a qubit. If we are restricted to a subspace of fixed fermion number parity, a topological qubit needs at least four Majorana zero modes, and hence two topological regions in the wire, separated by a non-topological region. This is implemented in our model with four boundary walls at the positions $\{ x_\text{wall}^{(1)}, x_\text{wall}^{(2)}, x_\text{wall}^{(3)}, x_\text{wall}^{(4)} \}$ (see Fig.~\ref{fig::setup}). The qubit space is then encoded in either the global even or odd parity subspace of the associated 4-fold degenerate ground state space $\{ |\text{GS}\rangle, \beta_0^\dagger |\text{GS}\rangle, \beta_1^\dagger|\text{GS}\rangle, \beta_0^\dagger\beta_1^\dagger|\text{GS}\rangle\}$. 
We will choose our logical qubit to be spanned by the ground states of the even-parity sector $\ket{\bar{0}} \equiv |\text{GS}\rangle$ and $\ket{\bar{1}} \equiv \beta_0^\dagger\beta_1^\dagger|\text{GS}\rangle$. We emphasize that, since the Hamiltonian $H(t)$ can be time-dependent, the spectrum $\epsilon_n(t)$, the normal mode operators $\beta_n(t)$, and the degenerate ground states $\{ \ket{\bar{0}(t)},  \ket{\bar{1}(t)} \}$ can also be time-dependent
\footnote{We note that the notation here could lead to the misconception that the qubit-space is identified with the occupation numbers of the Majorana zero modes alone. This would in turn imply that fidelity measures of the qubit-space can be computed by performing a partial trace over the bulk-eigenmodes.
In the following we will take the view that readout protocols that can distinguish fermionic and vacuum channels (see for example \cite{Aasen2016}) must, in some guise, project to the actual energy eigenstates and as such, should also be able to detect errors that come from states where the occupation numbers of some bulk modes, together with one or both Majorana zero modes, have been flipped.  In this paper we therefore do not see bulk excitations by themselves as an insurmountable problem. 
However we are concerned about the damage excitations can do when they, through a variety of mechanisms, find their way back to the ground state manifold.}.
In appendix~\ref{sect:BdG} we outline how excitations and superpositions thereof are handled within the BdG formalism. 

In what follows we will need to sometimes specify bulk eigenstates. In this case we will adopt the following notation 
\begin{align}
\ket{ 00 \{n_B \}} = \ket{0}_L \otimes \ket{0}_R \otimes \ket{n_B}  \non \\
\ket{ 11 \{n_B \}} = \ket{1}_L \otimes \ket{1}_R \otimes \ket{n_B}
\label{eq:estatenotation}
\end{align}
where $n_{B}$ is binary number encoding the bulk mode $\beta_n$ occupations and the $L/R$ subscript refers to the zero mode of the left/right wire. In this notion our two ground states would be written as 
$\ket{\bar{0}} \equiv \ket{ 00 \{0...0 \}}$ and $\ket{\bar{1}} \equiv \ket{ 11 \{0...0 \}}$

\subsection{Dynamics}
At some initial time $t = t_\text{init}$, we suppose that the system is in the state $\ket{\psi(t_\text{init})} = \alpha_0 \ket{\bar{0}(t_\text{init})} + \alpha_1 \ket{\bar{1}(t_\text{init})}$, where $\ket{\bar{0}(t_\text{init})}$ and $\ket{\bar{1}(t_\text{init})}$ are ground states of the Hamiltonian $H(t_\text{init})$ at that time. At a later time $t$, the system has evolved to the state $\ket{\psi(t)} = \mathcal{U}(t,t_\text{init}) \ket{\psi(t_\text{init})}$, where $\mathcal{U}(t,t_\text{init})=\mathcal{T}e^{-i\int_{t_\text{init}}^t H(t')dt'}$. If the boundary walls, and hence the Hamiltonian $H(t)$, are changed very slowly then, by the adiabatic theorem, in the ideal case, the system at a later time $t$ will remain in a superposition of the instantaneous ground states, $\ket{\psi_\text{ideal}(t)} = \alpha_0 \ket{\bar{0}(t)} + \alpha_1 \ket{\bar{1}(t)}$ with the same amplitudes as the initial state. This is the key feature of topological computation: if we assume adiabaticity, the only way to rotate within the ground state space is to ``braid" MBS around each other. We regard $\ket{\psi_\text{ideal}(t)}$ as the ideal, error-free evolution. Deviations from adiabaticity will lead to excitation of the bulk modes, corresponding to qubit loss. We quantify the qubit-loss by the probability $P_\text{loss}$ that the system is not in one of our instantaneous qubit states:
\begin{equation} P_\text{loss}(t) = 1 - \hspace{4pt} |\langle \psi(t) | \bar{0}(t)\rangle|^2 - |\langle \psi(t) | \bar{1}(t)\rangle|^2  \hspace{4pt}.
\end{equation}
This gives $P_\text{loss}=0$ if there is no qubit-loss and $P_\text{loss}=1$ if the qubit is completely lost. Since qubit loss corresponds to energy excitations, qubit-loss errors can, in principle, be detected by projective measurements of energy. More damaging are the undetectable errors resulting from bulk excitation returning to the wrong ground state. We consider two types of qubit error: bit flips and phase flips. A bit-flip error occurs if the system returns to the qubit space in the state $\ket{\psi_\text{bit}(t)} = \alpha_0 \ket{\bar{1}(t)} + \alpha_1 \ket{\bar{0}(t)}$ (i.e., with the qubit basis states exchanged $\ket{\bar{1}(t)} \leftrightarrow \ket{\bar{0}(t)}$). We quantify the bit-flip error with the probability: 
\begin{equation} P_\text{bit}(t) = |\langle\psi(t) | \psi_\text{bit}(t) \rangle |^2 . \end{equation} 
In the following study of bit-flip error we will generally choose $\ket{\psi(0)}=\ket{\bar{0}}$ and thus the bit flip error is simply given as $P_{\text{bit}} = |\braket{\psi(t)}{\bar{1} (t)}|^2$. 

Phase-flip error occurs if the system returns to the qubit state $\ket{\psi_\text{phase}(t)} = \alpha_0 \ket{\bar{0}(t)} - \alpha_1 \ket{\bar{1}(t)}$, with the probability: 
\begin{equation} P_\text{phase}(t) = |\langle\psi(t) | \psi_\text{phase}(t) \rangle |^2 . \end{equation} 
In the following discussions of phase-flip error we will choose $\ket{\psi(0)} = \ket{\psi_+(0)}\equiv \frac{1}{\sqrt{2}} (\ket{\bar{0}}+\ket{\bar{1}})$ and  $\ket{\psi_\text{phase}(t)} = \ket{\psi_-(t)} \equiv \frac{1}{\sqrt{2}} (\ket{\bar{0}(t)}-\ket{\bar{1}(t)})$.

All the above amplitudes are numerically evaluated using the Onishi formula\cite{Onishi1966,NuclearManyBody2004}. See Appendix \ref{sect:BdG} for more details.


\begin{figure*}[t]
\centering
\includegraphics[width=2\columnwidth]{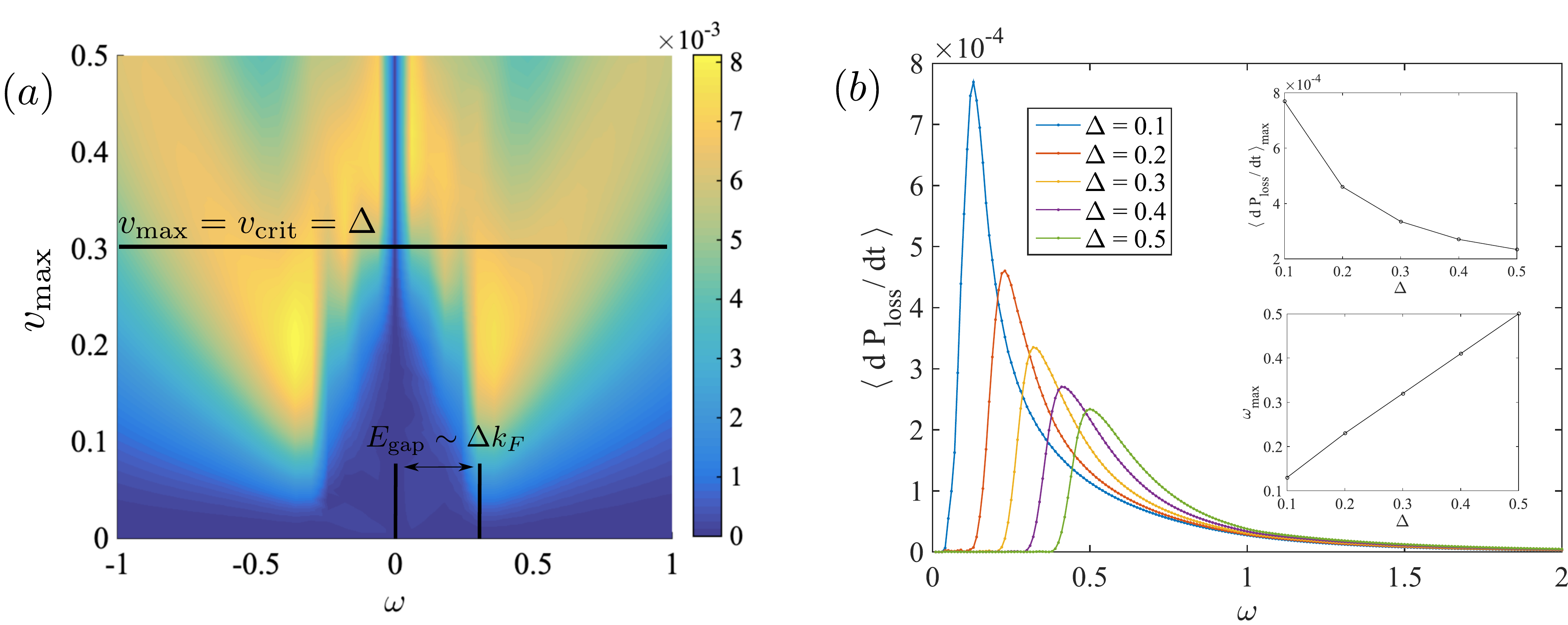}
\caption{(Colour online) (a) The time-averaged qubit-loss, as a function of the wall velocity parameters $v_{\text{max}}$ and $\omega$. The horizontal black line shows the critical velocity $v_{\text{crit}} = |\Delta| = 0.3$. To generate this figure we used $L=200, a=1, m=0.5, v(t) = v_{\text{max}} \sin(\omega t)$, and $\mu = 0.5$. The boundary potentials in this case were set to be essentially infinite ($\mu_{\text{boundary}}=-2000$) and the wall gradient with $\sigma = 1$ (see Appendix \ref{app::wall} for details on wall parameterisation). (b) A cross section of (a) at $ v_{\text{max}} = 0.02$. We see that the maximum qubit-loss rate occurs when the frequency is close to $E_{\text{gap}}$, but where the maximum value decays as $ 1/E_{\text{gap}}$.  The low frequency regime corresponds to the super-adiabatic limit and so we typically see very low rates of qubit loss error. For higher frequencies we see that error rate drops off as  $1/ \omega^4$. Crucially, because the maximum resonant frequency scales with $E_{\text{gap}}$, in the high frequency regime, increasing the topological order parameter can make the system more susceptible to errors.
}
\label{fig:OmegaQubitLoss}
\end{figure*}

\section{Qubit loss}
\label{sect:QubitLoss}

In this section we discuss how the principal source of qubit error (i.e. qubit loss) arises in a simple Majorana wire set-up with moving walls. First, in section \ref{sect:oscillating_Qubit_Loss}, we consider a wall that is oscillating sinusoidally in position. Such a motion may result, for example, from noise in the gates used to control the confining potential, or as a Fourier component of some deliberate wall motion. Then, in section \ref{sect:Adiabatic_Qubit_Loss} we consider a wall that is accelerated to some constant velocity and then, after some time, decelerated back to rest. This corresponds to a scenario where one needs to shuttle Majorana bound states between two different points in space.

\subsection{Qubit-loss due to an oscillating wall}
\label{sect:oscillating_Qubit_Loss}

We first consider the qubit loss $P_\text{loss}$ due to the motion of a single boundary wall, the wall at position $x_\text{wall}^{(1)}$, with an oscillating velocity profile of the form:
\begin{equation}
v^{(1)}_\text{wall}(t) = v_\text{max} \sin (\omega t).
\label{eq:v_wall_osc} 
\end{equation}
Since the wall is continuously moving, it is appropriate to calculate the time-averaged rate of qubit-loss \mbox{$\langle dP_\text{loss}/dt \rangle = \frac{1}{t'' - t'}\int_{t'}^{t''} dt (dP_\text{loss}(t)/dt)$}, over a time interval \mbox{$t^{''} - t^{'} = 2n \pi / \omega$} spanning some periods of the oscillation.  The colour map in Fig.~\ref{fig:OmegaQubitLoss}(a) shows $\langle dP_\text{loss}/dt \rangle$ as a function of the wall motion parameters $\omega$ and $v_\text{max}$. We can identify several features. First, when $v_\text{max} = 0$ or when $\omega = 0$ the qubit-loss rate is zero, as expected, since the wall is static in this case. For small values of $v_\text{max}$ the qubit-loss rate is small irrespective of the value of $\omega$. This is the adiabatic regime where the wavefunction of the system closely follows the ground state of the instantaneous Hamiltonian. However, for velocities larger than a critical velocity $v_\text{crit} = |\Delta|$ the qubit-loss rate is large, even for small frequencies $\omega$. This is consistent with the results of Ref. \onlinecite{Karzig2013,Scheurer2013}. In Fig.~\ref{fig:OmegaQubitLoss}(a) we see that the rate of qubit-loss is highest when the wall oscillation frequency $\omega$ is close to the gap energy $E_\text{gap} = \Delta k_F$. This is verified in Fig.~\ref{fig:OmegaQubitLoss}(b), where we plot a cross-section of 
Fig.~\ref{fig:OmegaQubitLoss}(a) at a fixed sub-critical value of $v_\text{max}$ for different values of the superconducting gap $\Delta$, and hence different values of the gap $E_\text{gap} = \Delta k_F$. Fig.~\ref{fig:OmegaQubitLoss}(b) also shows that for $\omega \gg E_\text{gap}$, the qubit-loss rate decreases quickly. We find numerically that this drop off is proportional to $1/\omega^4$.
\begin{figure*}[t]
\centering
\includegraphics[width=2\columnwidth]{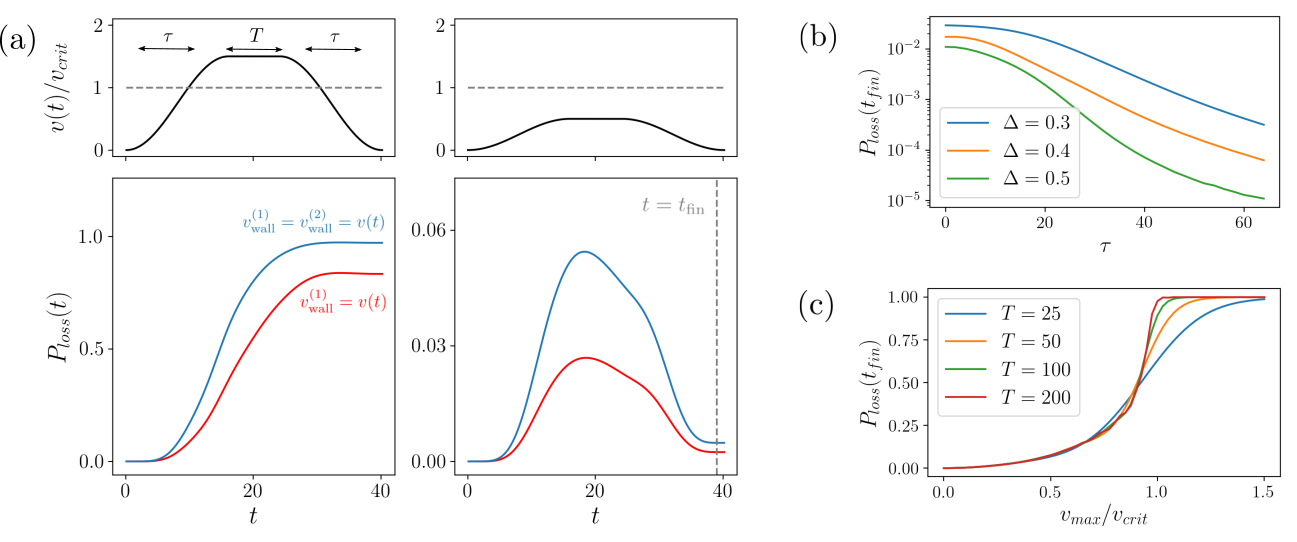}
\caption{ (Colour online) (a) Qubit-loss for $v_\text{max} > v_\text{crit}$ (left panel) and $v_\text{max} < v_\text{crit}$ (right panel). The corresponding wall velocity profiles are plotted above. We see that the final qubit loss is larger for two moving walls than for a single moving wall (blue lines vs. red lines) (Plotted for $\Delta = 0.4$). (b) The final qubit-loss at some time $t_\text{fin}$ after the walls have come to rest, plotted as a function of $\tau = 1/\omega$ (for $v_\text{max}=0.1$ and $T=30$). (c) The final qubit-loss in the limit of sudden wall acceleration ($\tau \to 0$). As $T$ gets large we see a sharp distinction emerge between the $v_\text{max}>v_\text{crit}$ and $v_\text{max}<v_\text{crit}$ regimes (here, $v_\text{crit} = |\Delta| = 0.2$). [Other parameters: in all figures, $a=0.5$, $m=0.5$, $L = 200$, $\mu = 1.0$ (in the topological region).]}
\label{fig:P_loss_MBS_transport}
\end{figure*}
\subsection{Qubit-loss in MBS transport}
\label{sect:Adiabatic_Qubit_Loss}

While the frequency $\omega_\text{max}$, at which we have maximum qubit-loss rate, grows with $\Delta$, the actual qubit-loss rate $\langle dP_\text{loss}/dt \rangle_{\text{max}}$ decreases [see insets of Fig.~\ref{fig:OmegaQubitLoss}(b)]. This decrease can be partially put down to the fact that the overall rate of qubit loss depends on the oscillation amplitude. For the choice of parameterisation given in Eq.~\eqref{eq:v_wall_osc} we have \mbox{$x^{(1)}_\text{wall}(t) = x^{(1)}_\text{wall}(0) + v_\text{max}[1 - \cos(\omega t)]/\omega$} and thus for fixed maximum velocity we have a smaller oscillation amplitude at higher frequencies. In similar calculations (not shown) where the wall movement is parameterised as $x(t) = x_0 + x_{\text{max}}\cos( \omega t)$, i.e., with a fixed amplitude $x_\text{max}$, we see the drop-off in qubit-loss rate for $\omega \gg E_\text{gap}$ scale as $1/ \omega^2$. 

For fixed frequencies larger than the gap [see e.g., $\omega \approx 0.6$ in Fig.~\ref{fig:OmegaQubitLoss}(b)] we see that the rate of qubit-loss increases as $\Delta$ is increased. This is in contrast to the situation in the low frequency adiabatic regime, where increasing $\Delta$ (hence increasing the gap $E_\text{gap}$) decreases the rate of qubit-loss. Thus while it makes sense to try to maximise the topological gap to counteract low-frequency noise and/or errors associated with deliberate motion of the topological boundary, the situation is more complicated if the range of perturbing frequencies extends above the frequency corresponding to the bulk gap.

\subsubsection{A single moving wall}

The oscillating velocity profile of Eq.~\ref{eq:v_wall_osc} is unsuitable if the goal is to transport a MBS between two different positions on the wire. To model transport, we consider a velocity profile of the form:

\begin{equation}
v_\text{wall}^{(1)}(t) = \begin{cases}
v_\text{max}\frac{1-\cos(\omega t)}{2} & 0 \leq t \leq \frac{\pi}{\omega} \\
v_\text{max} & \frac{\pi}{\omega} \leq t \leq \frac{\pi}{\omega} + T \\
v_\text{max}\frac{1-\cos(\omega t - \omega T)}{2} & \frac{\pi}{\omega} + T \leq t \leq \frac{2\pi}{\omega} + T  \\ 
0 & \text{otherwise}.
\end{cases}
\label{eq:v_wall_1} 
\end{equation}

\noindent This results in an acceleration from zero velocity to a maximum velocity $v_\text{max}$, followed by a period $T$ at constant velocity, and finally a deceleration back to zero velocity [see Fig.~\ref{fig:P_loss_MBS_transport}(a), top panels]. A small frequency $\omega$ corresponds to a slow acceleration of the wall, while a large value corresponds to sudden acceleration.

The qubit-loss due to a single moving wall for MBS transport has been studied in several previous works\cite{Karzig2013,Scheurer2013,Bauer2018}, where it was found that there is a critical velocity $v_\text{crit} = |\Delta|$, below which qubit-loss can be avoided for sufficiently slow acceleration of the wall. Above the critical velocity, qubit-loss will occur even if the acceleration is small \cite{Scheurer2013}. The difference between the $v < v_\text{crit}$ regime and the $v > v_\text{crit}$ regime is revealed by comparing the left and right panels in Fig.~\ref{fig:P_loss_MBS_transport}(a). We see that at some time $t_\text{fin}$ after the wall movement has ended, the final qubit-loss $P_\text{loss}(t_\text{fin})$ is small if $v_\text{max} < v_\text{crit}$, but can be very large if $v_\text{max} > v_\text{crit}$. 

We note that in the $v < v_\text{crit}$ regime, the motion is by no means adiabatic (except at very low velocity) and there is considerable qubit loss during the wall's motion in the acceleration phase. However the system returns to the ground state manifold as the wall is slowed down again. This means the wall can be moved far faster than what would be naively considered adiabatic, without permanent qubit loss. This has also been noted in earlier works on this subject\cite{Karzig2013,Scheurer2013} and it is linked to the notion of super-adiabaticity, where the motion can be considered adiabatic in a moving frame  (see e.g Ref. \onlinecite{Berry1987}). We explore this further in Appendices \ref{sect:CriticalVelocityAppendix} and \ref{sect:EffectiveWallMass} where we note that it is possible to think of  the instantaneous energy increase as a kinetic energy of the boundary wall. The final excess energy after the wall has stopped is a small correction to this behaviour, vanishing exponentially with increasing $\Delta$ and $\tau$.  

\begin{figure*}[t!]
\includegraphics[width=2.2\columnwidth]{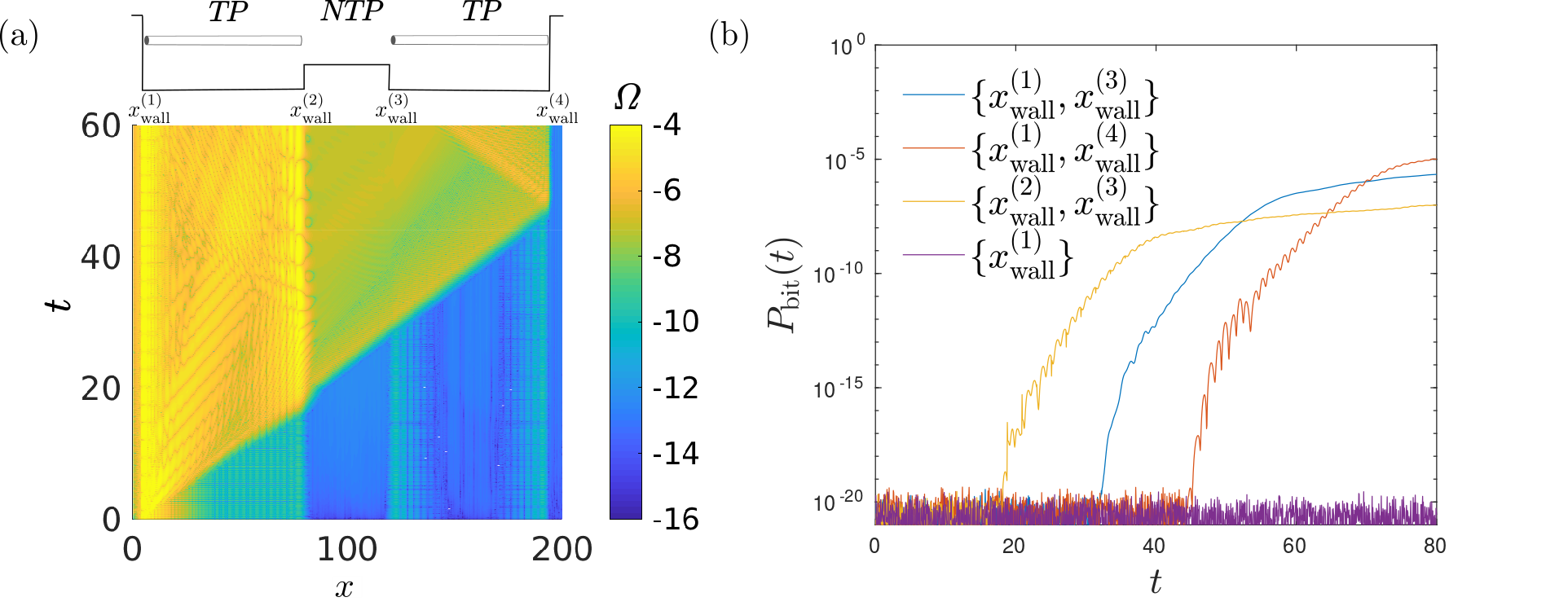}

\caption{(Colour online) (a) The density wave packet generated by a \emph{single} moving wall at $x_\text{wall}^{(1)}$ travels at a velocity approximately equal to $v_F$. We can see the wave tunnelling through the non-topological barrier separating the topological phases (barrier height $\mu_\text{barrier} = -2.5$). Here, $v_{\text{max}}=0.3$, and $\omega =3$. (b) The bit-flip error $P_\text{bit}$ increases suddenly as the excitation produced by the movement of one wall hits another moving wall (legend shows which walls are in motion). The time at which the error begins to increase can be estimated as approximately $t \approx |x_\text{wall}^{(i)} - x_\text{wall}^{(j)}| / v_F$ where $|x_\text{wall}^{(i)} - x_\text{wall}^{(j)}|$ is the distance between the two moving walls. [Other parameters for both figures: $m=0.5$, $a=0.5$, $\Delta = 0.4$, $L=200$, and $\mu = 1.0$ (in the topological region).]}
\label{fig:bit_flip_error}
\end{figure*}

To see the dependence of the qubit-loss on the wall acceleration, in Fig.~\ref{fig:P_loss_MBS_transport}(b) we plot the final qubit-loss $P_\text{loss}(t_\text{fin})$ as a function of $\tau = 1/\omega$ in the subcritical $v<v_\text{crit}$ regime. For large $\tau$, corresponding to slow wall acceleration, we see that the qubit-loss follows an exponential decay (in agreement with the results of Ref. \onlinecite{Scheurer2013}). However, the behaviour is different in the region where $\tau \to 0$, corresponding to sudden movement of the wall. In this limit, the qubit-loss reaches its maximum value (for a given $v_\text{max} < v_\text{crit}$). Nevertheless this maximum qubit-loss is usually small and decays with increasing $\Delta$ and we conclude that there is no critical acceleration analogous to the critical velocity in this set-up.

In Fig.~\ref{fig:P_loss_MBS_transport}(c) we plot the qubit-loss as a function of $v_\text{max}$ when $\tau \to 0$ (i.e., in the limit of sudden acceleration). As the time $T$ spent at at the maximum velocity increases we see a sharp distinction emerge between the $v_\text{max} > v_\text{crit}$ and $v_\text{max} < v_\text{crit}$ regimes: for \mbox{$v_\text{max}>v_\text{crit}$} the qubit-loss approaches unity and all theinformation is lost into bulk states, while for $v_\text{max}<v_\text{crit}$ we have $P_\text{loss}(t_\text{fin}) < 1$. This provides further evidence for the interpretation of $v_\text{crit} = |\Delta|$ as a critical velocity.

\subsubsection{Correlated movements and effective wall mass}

We also consider the motion of two boundary walls (the two leftmost walls at $x_\text{wall}^{(1)}$ and $x_\text{wall}^{(2)}$), both by identical velocity profiles $v_\text{wall}^{(1)}(t) = v_\text{wall}^{(2)}(t)$ given by Eq.~\ref{eq:v_wall_1}, so that the two walls are moving in sync and the length of the wire remains constant, see e.g.~Ref.~\onlinecite{Scheurer2013}.

The numerical results shown in Fig.~\ref{fig:P_loss_MBS_transport}(a) show that there is no advantage to coordinated wall movement, compared to a single moving wall. Indeed in the right-panel of Fig.~\ref{fig:P_loss_MBS_transport}(a) we see that for $v_\text{max}<v_\text{crit}$ the qubit-loss is approximately a factor of two higher in this case than for a single moving boundary wall, and that this factor of two persists throughout the movement protocol. In appendix \ref{sect:EffectiveWallMass} we provide a further discussion of the same phenomena where, instead of qubit-loss, we examine the energy above the instantaneous ground state. We find that, up to small exponential corrections due to changes in velocity, the energy increase scales as $\frac{1}{2} M v^2$ where the effective mass scales as  $M \propto p_F^2/E_{\text{gap}}$. 

\section{Undetectable qubit errors}
\label{sect:secodaryerrors}

\subsection{Bit-flip error}
\label{sect:BitflipError}
In section \ref{sect:QubitLoss} we saw that the movement of the boundary wall can lead to qubit-loss that can be recovered, to some extent, when $v<v_{\text{crit}}$. This raises the question: can the system return to the qubit space with an error? In this section we will show that bit-flip errors can indeed occur, and we will explore how they are related to excitations produced during the wall movement. 

First, we note that, in general, wall movement generates local excitations because of the local nature of the perturbation. These local excitations can then propagate through the wire over time. To investigate this, we consider a single wall (at the position $x_\text{wall}^{(1)}$) oscillating sinusoidally as in Eq.~(\ref{eq:v_wall_osc}) [although similar results can be obtained with the shuttling movement protocol given by Eq.~(\ref{eq:v_wall_1})]. In Fig.~\ref{fig:bit_flip_error}(a) we plot the deviations from the ideal particle density: 
\begin{equation} 
{\Omega_{j}}(t) = \bra{\psi(t)} c_j^\dagger c_j \ket{\psi(t)} - \bra{\psi_\text{ideal}(t)} c_j^\dagger c_j \ket{\psi_\text{ideal}(t)}, \label{eq::mode} 
\end{equation} 
as a function of time $t$ and the position $x=ja$ along the wire, due to the motion of the boundary wall. The plot clearly shows a particle density wave that propagates across the wire. The velocity of the propagating wave can be read off from the plot and is found to be approximately equal to the Fermi velocity $v_F$. This value for the velocity is essentially due to the fact that if we consider the system in the moving frame, excitations, made out of quasiparticle and quasiholes with relative momentum $2k_F$, are favoured energetically (see Appendix \ref{sect:CriticalVelocityAppendix} for more details).
\begin{figure}[t]
\centering
\includegraphics[scale=0.2]{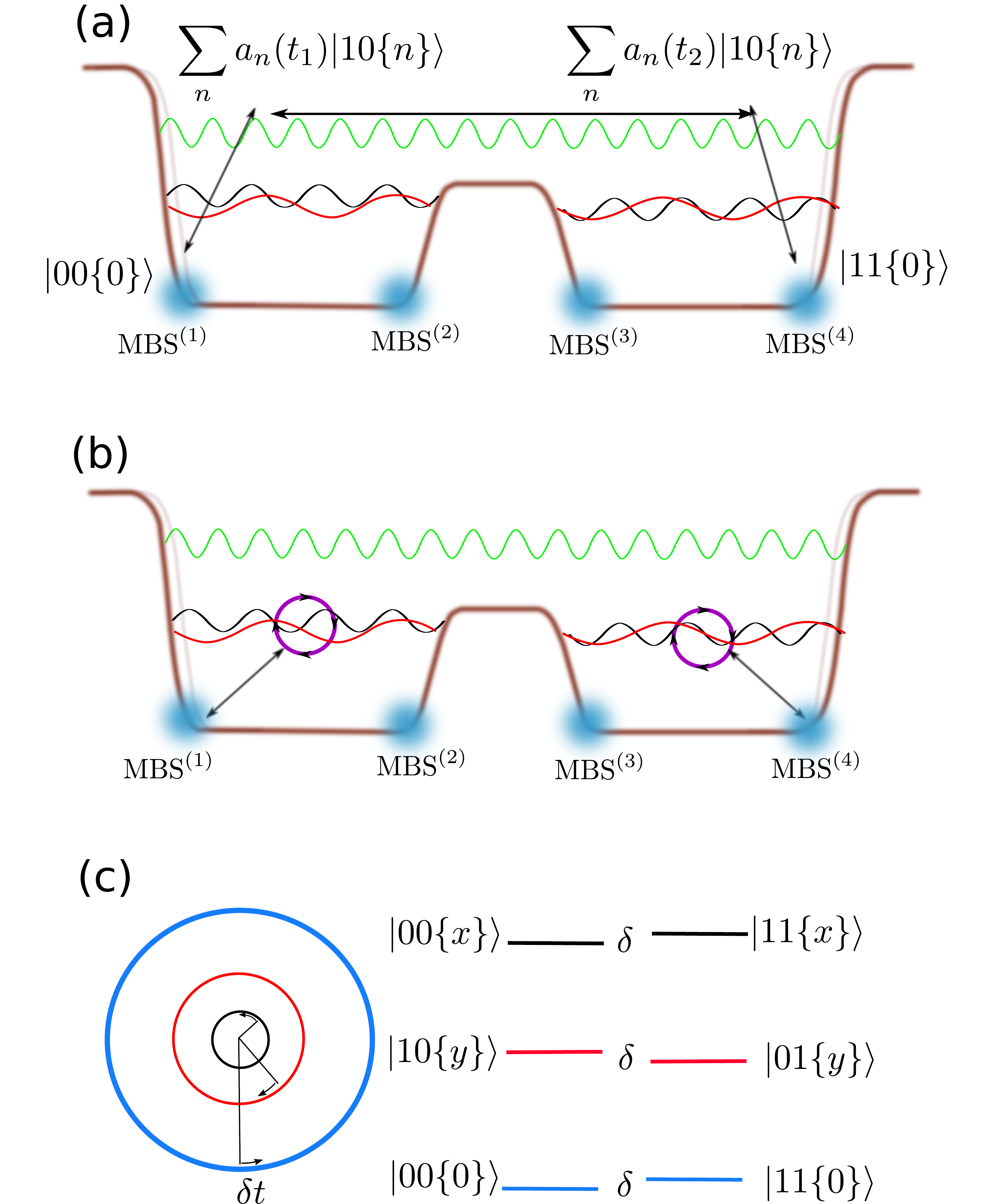}
\caption{ (Colour online) (a) Schematic showing how excitations originating at the outer walls can dynamically evolve through the system and cause a bit-flip error ( $\ket{\bar{0}} \leftrightarrow \ket{\bar{1}}$ (b) For the anomalous phase rotation the moving boundary wall generates population transfer between ground states $\ket{\psi_{\pm}}$ and excited states. Energy mismatches between states in the bulk can then drive a rotation that either reinforces or opposes the natural rotation due to the ground state splitting. (c) Schematic of how the zero-mode splitting affects ground and bulk-states (see Eq.~\ref{eq:estatenotation} for explanation of the notation).  Deviations from the natural phase rotation (blue) can occur because some excited states (red) have a phase rotation that acts in the opposite direction. The lowest lying pairs of excited states will have a single fermion transferred from a Majorana mode (here the left edge mode) to an excited bulk mode, which means the splitting in these pairs is the same as the splitting between the Majorana modes in the odd fermionic parity sector, which is opposite to the splitting in the even parity sector which contains the qubit states.  Higher energy states (black) can have the same occupation of the Majorana modes as the ground state and then do not contribute to the phase error as they are split in precisely the same way as the ground states.}
\label{fig:qubiterrors}
\end{figure}

We can now understand a mechanism for the occurrence of bit-flip errors. The excitations generated at a moving wall can travel across the wire, and tunnel through non-topological regions in the wire, carrying information between MBS. The resulting interaction between MBS can, in principle, induce bit-flip errors in the topological qubit. This mechanism is illustrated schematically in Fig.~\ref{fig:qubiterrors}(a). It is important to note that, in order for the MBS to interact with the incoming density wave packet, one of the walls on the other side of the system must also be in motion. If this is not the case, the density wave packet will not be able to dissipate at the wall: it will simply be reflected or transmitted without any chance for the excitation to decay back into the degenerate ground state.

This is verified numerically in Fig.~\ref{fig:bit_flip_error}(b), where we show the bit-flip error $P_\text{bit}$ due to the oscillation of boundary walls (the walls that are moving are indicated in curly brackets in the legend). We see that there is no bit-flip error if a single wall is moved and all other walls remain static. However, if two walls are moved, bit-flip errors $P_\text{bit}$ appear abruptly after some delay. The times at which the bit-flip errors begin to appear can be estimated as the time taken for the propagating wave to reach the other moving wall, i.e., as $t \approx |x_\text{wall}^{(i)} - x_\text{wall}^{(j)}| / v_F$ where $|x_\text{wall}^{(i)} - x_\text{wall}^{(j)}|$ is the distance between the two moving walls. This provides strong evidence that it is the stray excitations propagating along the wire that are responsible for the measured bit-flip error. 

\subsubsection{Effects of disorder}\label{sect:Disorder}

\begin{figure}[t!]
\centering
	\includegraphics[scale=0.615]{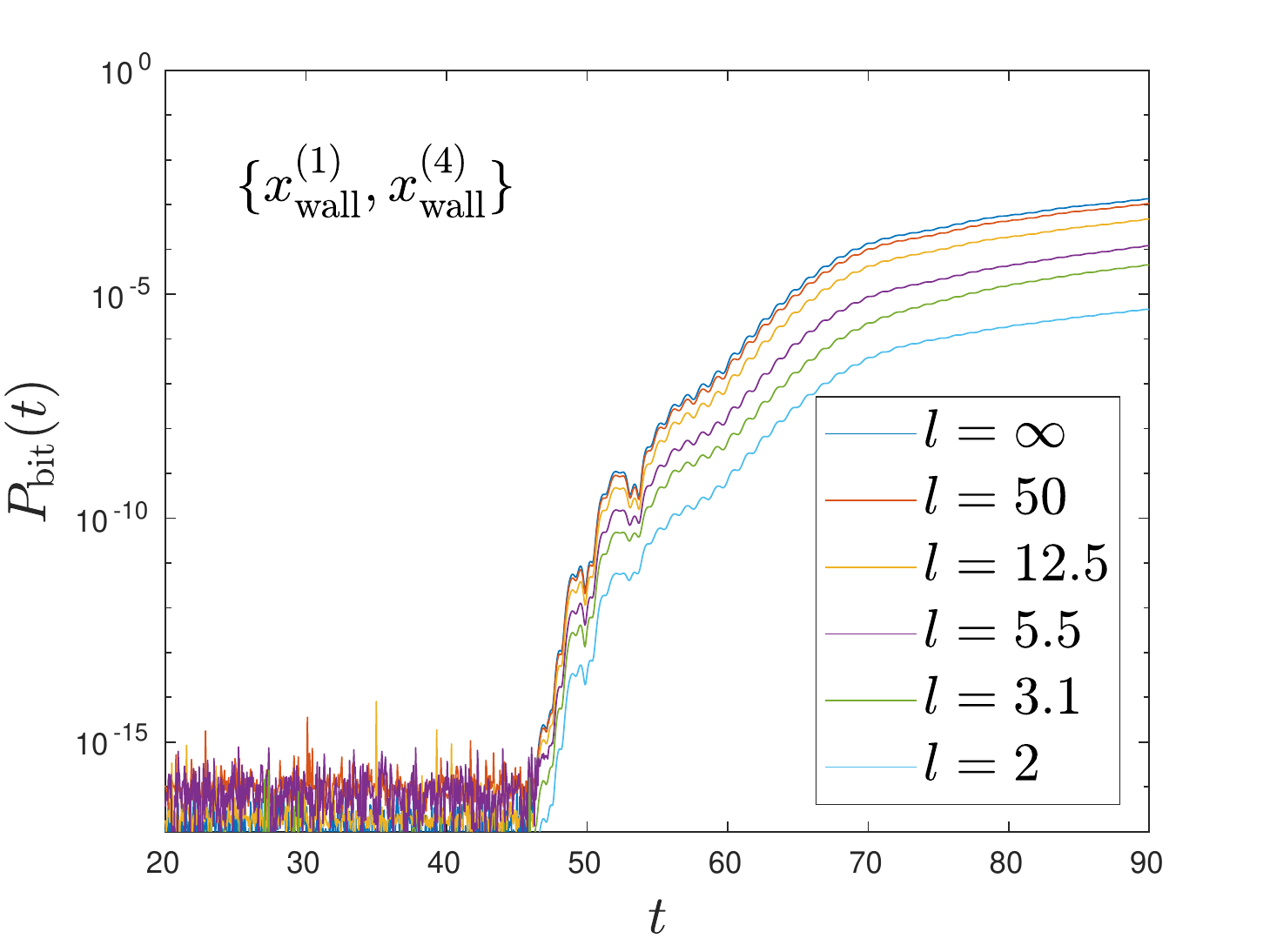}
	\caption{ (Colour online) Bit-flip error is suppressed by adding disorder in the middle of the left-hand wire. Here, the disordered region is of length $\sim 40$, and a shorter localisation length $l$ corresponds to increased disorder. We use $v_\text{max}/v_{\text{crit}} = 0.75$, the middle barrier height is $\mu_\text{barrier} = -2.5$, $\sigma = 2$ and the data is averaged over 20 disorder realizations. [Other parameters used for the figure: $m=0.5$, $a=0.5$, $\Delta = 0.4$, $L=200$, and $\mu = 1.0$ (in the topological region).]}
\label{fig::DisorderScan}
\end{figure}

The central result of the previous section is that bit-flip error can occur in a two wire setup if there is a process whereby a quasi-particle is excited in one wire, tunnels through the barrier to the other wire, and then decays back to the ground state manifold. In order to avoid such a process, one straightforward approach would be to increase the barrier between the wires to prevent inter-wire tunneling. However, within the schemes to manipulate quantum information using wire networks, see for example Ref. \onlinecite{Alicea2011}, one often needs to bring neighboring Majorana modes together, a process which would render the barrier more transparent and which would potentially also excite more propagating quasi-particles. In this section we show that another solution to prevent quasi-particle propagation is to introduce some disorder into the wires themselves. 

Naively, we might introduce disorder throughout the wire. However there are downsides to doing this. Firstly, in static scenarios, disorder decreases the gap between the ground state and the bulk excitation spectrum, increasing the bound-state decay length \cite{Brouwer2011,Brouwer2011b,DeGottardi2013} and results in a topological space that is less resilient to qubit-loss. Moreover it has been shown \cite{Karzig2013} that moving the confining potential over a disordered region results in a dramatically lower critical velocity, thus severely hampering the rate at which gates can be mechanically performed. 

An effective compromise is to deliberately disorder only the central wire regions. This allows for an effective critical velocity of the wire ends that is equal to the $p$-wave order parameter $\Delta$ while also providing a region where bulk excitations are prevented from propagating, reducing the probability that the excitations originating at one wall can tunnel to another wire. This approach still allows for low barrier transparency and so the original schemes\cite{Alicea2011} for braiding and rotating Majorana pairs can be performed as usual. 

In Fig.~\ref{fig::DisorderScan} we present the results of our numerical simulations, where disorder has been introduced in the central regions of both wires.  As earlier, we simulate a scenario where there is an oscillation of two walls $x_{\text{wall}}^{(1)}$ and $x_{\text{wall}}^{(4)}$ in different wires. The figure shows the decrease in probability of bit-flip error as we decrease the localization length $l$ in the disordered region. This indicates that as the disorder is increased, the wave associated with excitations accrued at the boundaries cannot propagate through the wire, and hence the junction, to induce a bit-flip error. 

\subsection{Phase corrections in a single wire}
\label{sect:phase_error}

In a two wire setup, phase-flip error can occur in precisely the same way as bit-flip error: excitations originating in one wire tunnel into the other wire and relax to the ground state.  However there is another process through which phase error can occur, which importantly does not involve tunneling between wires. This process is shown schematically in Figs. \ref{fig:qubiterrors}(b) and \ref{fig:qubiterrors}(c). The mechanism relies on the fact that splittings between states in the bulk do not necessarily correspond with the splitting that occurs in the ground-state. In systems that experience continual qubit-loss this can lead to a small but systematic phase error being returned to the qubit space. 
 
A necessary feature here is that pairs of bulk excitations that differ in the occupation of the edge modes have slightly different energies. In non-interacting systems this can only happen if there is also some small splitting in the ground state leading to a natural dynamical oscillation between the $\ket{\psi_{\pm}} = \frac{1}{\sqrt{2}} (\ket{\bar{0}}\pm \ket{\bar{1}})$ states. This rate of oscillation depends on the splittings due to the near zero-modes in both wires such that:
\begin{equation}
\label{eq:naturalrotation}
|\braket{\psi_-}{\psi_+(t)}| =  \sin\left(\frac{\delta (t) }{2}\right) 
\end{equation}
with $\delta(t) = \int_0^{t} [E_{\bar{1}}(t') - E_{\bar{0}}(t')] dt'$ depends on the difference in energy between the ground states, and hence is related to the decay length of the Majorana edge states.  In Ref. \onlinecite{Scheurer2013} it was noted that because the decay length of the bound states take on a relativistic like correction in moving frames, this rate of rotation within the ground state space can in addition depend on velocity of the system.  

The correction to the phase oscillation that we study here  depends on there being a continual qubit-loss/gain channel between the ground state and excited states. If this is the case then energy mismatches between bulk eigenstate pairs will lead to a systematic deviation from this ``natural" Rabi oscillation.  The mechanism is discussed in more detail in Figs. \ref{fig:qubiterrors}(b) and (c).
\begin{figure}[t]
\includegraphics[scale=0.95]{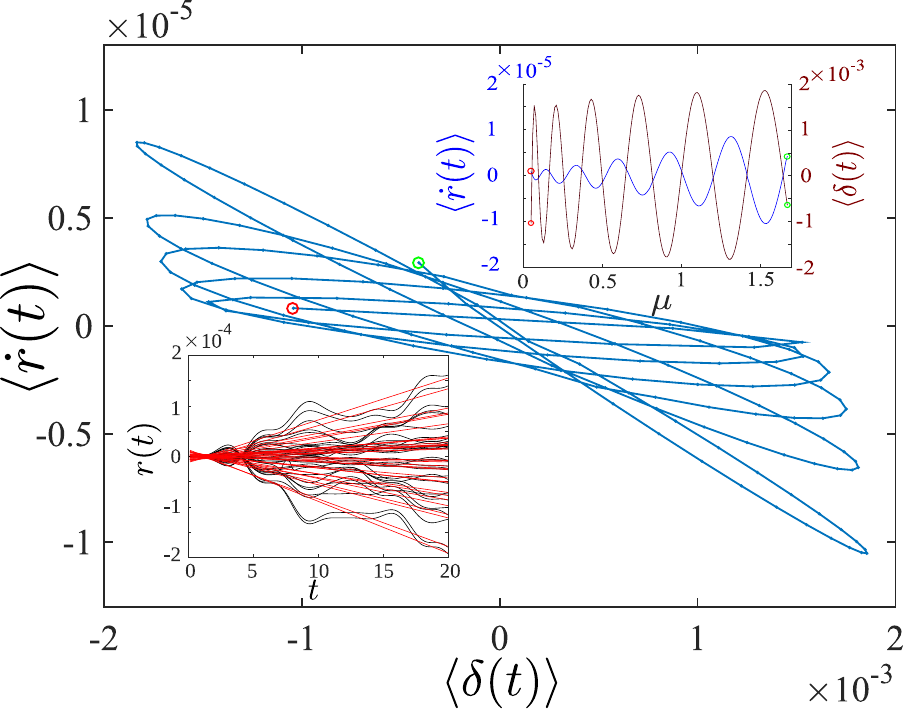}
\caption{
(Colour online) Main figure and upper inset: Qubit-loss induced deviation from the natural phase oscillation tends to act in the opposite direction to the natural oscillation. The red dots indicate $\mu=.03$ and the green $\mu=1.68$.  Lower inset: The deviation $r(t) =\phi(t)-\delta(t)$ (black) and the slopes of the fitted red lines give $\langle \dot{r} (t) \rangle$.  To generate this figure two wires of lengths $45$ and $40$ were used with an essentially infinite barrier between them. The continuum order parameter was set to $\Delta = 0.4$ in the left wire and $\Delta=0.8$ in the right.  The effective electron mass was set to $m=0.5$ and a lattice constant of $a=0.5$ was used.  The velocity of of the left hand boundary of the left wire was moved according to $v(t)=0.1 \sin( 2 t)$}
\label{fig:phaseresults}
\end{figure}

To analyze the deviation from the natural rotation in Eq.~(\ref{eq:naturalrotation}) we first recall that throughout this manuscript we take the view that parts of a state that leave the ground state space are detectable, and therefore we are interested only in the phase rotation within the ground-state manifold.  We therefore need to calculate the conditional probability that the state $\ket{\psi(t)}$ (where $\ket{\psi(0)} = \ket{\psi_+(0)}$) is measured in the instantaneous ground-state $\ket{\psi_\text{phase}}=\ket{\psi_-(t)}$, given that time-evolved state is projected to the ground state manifold.  This probability is given as 
\begin{equation}
R(t) =  {\frac{P_{\text{phase}} (t)}{ 1 - P_{\text{loss}}}}= \frac{|\braket{\psi(t)}{\psi_-(t)}|^2}{|\braket{\psi(t)}{\psi_-(t)}|^2+|\braket{\psi(t)}{\psi_+(t)}|^2}.
\end{equation}
From here, we can calculate the phase rotation angle within the ground-state manifold, giving
\begin{equation}
\sin\left(\frac{\phi(t)}{2}\right)=\sqrt{R(t)},
\end{equation}
and then the difference from the natural rotation angle $\delta(t)$ as
\begin{eqnarray}
r(t) &=& \phi(t) - \delta(t)  
\end{eqnarray}

In Figure \ref{fig:phaseresults} we plot the behaviour of $dr/dt$ as a function of the chemical potential $\mu$  and the natural phase oscillation due to the small even splitting in both wires. We see that the rate of this phase deviation opposes the mean splitting between the even-odd pairs $\langle \delta (t) \rangle$ that drives the natural phase oscillation. 
The mechanism that we describe here relies on there being differences in energies between bulk even-odd eigen-pairs. In a non-interacting system this is necessarily due to a splitting of the edge modes (making them no longer perfect zero modes). This then also implies that there is a splitting between ground states and, as a result, a natural non-topological rotation within the qubit space.  This process can be counteracted by making the wire longer, as this will reduce the overlap between the edge modes.
The situation is different  in interacting systems because in the presence of interaction, the many particle energy spectrum can no longer be described in terms of single particle modes. One thus expects  that there may not be an even-odd degeneracy for bulk states when interactions are strong, even though the ground state degeneracy is robust due to its topological nature. Indeed there is significant evidence that the even-odd symmetry within higher energy eigen-pairs breaks down in regions where bands with different fermion numbers are energetically similar \cite{Gangadharaiah2011,Goldstein2012, McGinely2017,Miao2017,Kells2015a,Kells2015b}.
It may therefore be possible that phase errors similar to the one described here are possible in the presence of non-adiabaticity and interactions, even though the ground state is essentially degenerate. 

\section{Concluding remarks}
\label{sect:Conclusions}
In this paper we have investigated possible sources of errors in a toy model of a topological qubit. Our motivation was to understand how error process can arise due to non-adiabatic variation of system parameters. As such, our analysis excludes errors due to interaction with an external environment, and therefore the results only incorporate a subset of possible error processes. Likewise, our analysis does not attempt to incorporate details of specific physical realisations of Majorana based qubits. However, the simplicity of the $p$-wave model should make it relevant to some degree to more realistic device designs (e.g. tetron and hexon proposals \cite{Plugge2017,Karzig2017}) and scalable architectures built from them\cite{Litinski2018}.

As we have stated throughout the paper, we take the viewpoint that the topological qubit consists of two ground states of the Hamiltonian. This is subtly different to the  alternative view where one understands the qubit as the reduced state of the zero-modes obtained by tracing over all excited modes.  Our perspective aligns with the original schemes for 2D topological computers in which measurement of the fusion channels occur as a projections to the many-body ground-states of the system. As we have seen this perspective implies that the system is, at least initially, protected from local errors.  

In this context then then our results specifically show how errors can arise in an isolated topological memory, assuming we have a perfect readout protocol. While this approach may not fully reflect the state of current devices, it is what allows us to resolve how errors propagate in time and to see how initial qubit-loss error can eventually lead to more damaging errors. These errors, even with perfect resolution of the fusion channels, cannot be detected without some conventional error correction protocols and the associated computational overhead. For bit-flip error we have shown that this must necessarily involve some tunneling from one wire to another, but for phase error we demonstrated how this could in principle happen via some local process within one wire. For non-interacting models we argue that this results in a correction to qubit rotation due to finite size splitting, in addition to that due to moving reference frames \cite{Scheurer2013}. In follow up work we aim to assess whether this effect can also be triggered by electron-electron interactions \cite{Coopmans2019}.

\vspace{15pt}

\begin{acknowledgments}
The authors would like to thank Luuk Coopmans, Kevin Kavanagh, Niall Moran, Stephen Nulty, Falko Pientka and  Alessandro Romito for useful discussions and comments.  A.C., D.P. and J.K.S. acknowledge financial support from Science Foundation Ireland through Principal Investigator Awards 12/IA/1697 and 16/IA/4524. A.C. was supported through IRC Government of Ireland Postgraduate Scholarship GOIPG/2016/722. S.D. and G.K. acknowledge support from Science Foundation Ireland through Career Development Award 15/CDA/3240. 
\end{acknowledgments}

\appendix

\section{Bogoliubov de Gennes formalism and ground state overlaps}
\label{sect:BdG}
In this section we provide a brief review of the BdG formalism applied to a $p$-wave superconductor. For a more detailed treatment of this material see Ref. \onlinecite{NuclearManyBody2004}. 

The Hamiltonian in Eq.~(\ref{eq:Hamiltonian}) can be conveniently written as $H(t) = \Psi^\dagger H_\text{BdG}(t) \Psi$, where $\Psi^\dagger = (c_1^\dagger, \hdots, c_N^\dagger, c_1, \hdots, \: c_N)$ is a row vector of fermionic creation and annihilation operators. The instantaneous Bogoliubov-de Gennes (BdG) Hamiltonian $H_\text{BdG}(t)$ can be diagonalised: $W^\dagger H_\text{BdG}W = \text{diag}(\epsilon_0,...,\epsilon_{N-1}, -\epsilon_0,...,-\epsilon_{N-1})$, by a unitary transformation of the form: 
\be \label{Wmatrix}
W(t) = \left( \begin{array}{cc} U(t) & V^*(t) \\ V(t) & U^*(t) \end{array}  \right) ,
\ee
where $U$ and $V$ are $N \times N$ complex matrices. This unitary also leads to the transformation of the fermionic modes given in Eq.~\ref{eq::bogoliubov} so that the Hamiltonian is written as 
\begin{equation}
H = \sum_{n}\epsilon_n \left(\beta_{n}^{\dag}\beta_{n} - \frac{1}{2}\right).
\end{equation}
For the translationally invariant system with periodic boundary conditions, the bulk single-particle spectrum can be given in terms of momentum as
\begin{align*}
\epsilon_{n} &= \sqrt{ (-\tilde{\mu} - 2 w \cos(k_n a))^2 + 4\tilde{\Delta}^2 \sin^2 (k_n a) } ,
\end{align*}
with $k_n = 2\pi n / Na$ in the first Brillouin zone. With our setup we identify the Dirac fermion zero modes of the left wire and of the right wire as:
\begin{align}
\beta_L = \frac{1}{2}( \beta_0 +i \beta_1) ,  \quad \text{and} \quad \beta_R = \frac{1}{2} (\beta_2+ i  \beta_3) .
\end{align}
with $\beta_n$ defined as in Eq.~(\ref{eq::bogoliubov}) in the main text.

All the probabilities in our simulations are obtained using the Onishi formula\cite{Onishi1966,NuclearManyBody2004}:
\begin{equation}
    |\langle{\psi(t) }|{\psi'(t)}\rangle|^2=  \text{det}(U(t)^* U'(t) + V(t)^* V'(t)) ,
    \label{eq:Onishi}
\end{equation}
where $U'$ and $V'$ are defined as:
\begin{equation}
\begin{pmatrix}U'(t) \\ V'(t)\end{pmatrix}=\mathcal{U}_\text{BdG}(t)\begin{pmatrix}U(0) \\ V(0)\end{pmatrix} ,
\end{equation} where $\mathcal{U}_\text{BdG}(t) = \mathcal{T}\exp\left( \int_0^t dt' H_\text{BdG}(t') \right)$.

On a practical level, to consider an excitation of the system, one can think of swapping our definition of $\beta_n$ with that of $\beta_n^\dagger$.  This corresponds to a swap of the $n^{\text{th}}$ column of $W$ with the $N+n^{\text{th}}$.  The time evolved state overlap with different ground states can then be obtained by swapping the columns $U_{jn} (V_{jn})$ with $V_{jn}^* (U_{jn}^*)$  \cite{NuclearManyBody2004}. For example in order to make the $\ket{\bar{1}}$ state we would swap columns as $1 \leftrightarrow N+1$ and $2 \leftrightarrow N+2$.  The same trick also works to make ground state superpositions  $ \propto \ket{\bar{0}} \pm \ket{\bar{1}} $. In this case one rotates within the relevant $\beta_n$subspace:
\begin{align}
& \beta_0^\dagger \rightarrow \frac{1}{2} ( \beta_0^\dagger \pm \beta_1) \\
& \beta_1^\dagger \rightarrow \frac{1}{2} ( \beta_1^\dagger \mp \beta_0) \non
\end{align}

\begin{figure}[t]
\centering

	\includegraphics[scale=0.5]{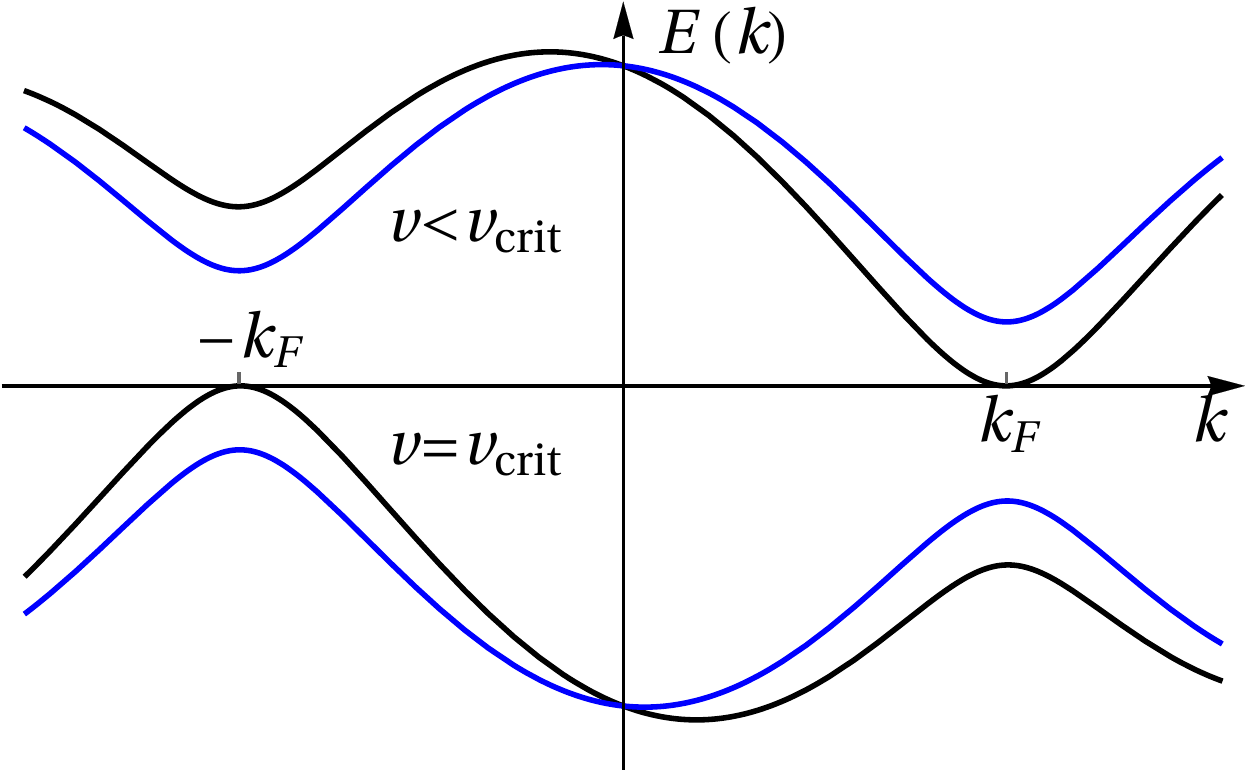}
\caption{
Bulk single particle energies of the instantaneous Hamiltonian in the ``moving frame'' for $v<v_{crit}$ and $v=v_{crit}$, where the gap is closed.}
\label{fig::BandsCriticalVelocity}
\end{figure}

\section{Boundary wall}\label{app::wall}

We model a boundary wall separating a topological and non-topological region by a sigmoid function for the chemical potential: \begin{equation} \mu (x) = \mu_{-} + \frac{\mu_{+} - \mu_{-}}{1 + e^{-(x - x_\text{wall})/\sigma}} , \label{eq:boundary_wall} \end{equation} where $x_\text{wall}$ is the position of the wall, $\sigma$ characterises the steepness of the wall, and $\mu_{\pm}$ is the chemical potential in the limit $x - x_\text{wall} \to \pm\infty$ (see Fig.~\ref{fig::setup} for an illustration). For convenience, in Eq.~\ref{eq:boundary_wall} we use the continuous label $x$ instead of the discrete label $j$ [i.e., $\tilde{\mu}_j \to \mu(x)$], although the lattice sites always appear at the discrete positions $x = ja$, where $a$ is the lattice spacing and $j=1,...,N$ is an integer.

Multiple walls along the wire are modelled by adding sigmoid functions at various positions ($x_\text{wall}^{(2)}$, $x_\text{wall}^{(3)}$, etc.) along the wire, with an appropriate normalisation, to give the desired chemical potential between any two walls. In the two-wire scenario described in section \ref{sec:topo_qubit} and in Fig.~\ref{fig::setup}, we choose the normalisation such that the chemical potential at the outer boundary walls is $\mu = 20w$, and at the middle wall is $\mu = 2.5w$ (unless otherwise stated). 

\section{The critical velocity and propagating excitations}
\label{sect:CriticalVelocityAppendix}
In this model the existence of a critical velocity $v_\text{crit}$ was first pointed out in Ref.~\onlinecite{Scheurer2013}. It it best understood in the continuum limit after transforming to a frame of reference that moves at the velocity $v(t)$ of the boundary wall. The transformation is implemented by the unitary operator $\mathcal{W}(t)=\exp\{-iP\int_0^t v(t')dt'\}$, where $P$ is the momentum operator. The Hamiltonian in the moving frame is:
\begin{eqnarray}
    H' &=& \mathcal{W}(t)H(t)\mathcal{W}(t)^\dagger+i\frac{d\mathcal{W}(t)}{dt}\mathcal{W}^\dagger(t) \nonumber \\
       &=& H(0)+P v(t) .
\end{eqnarray}
\begin{figure}[t!]
\centering
\includegraphics[scale=0.3]{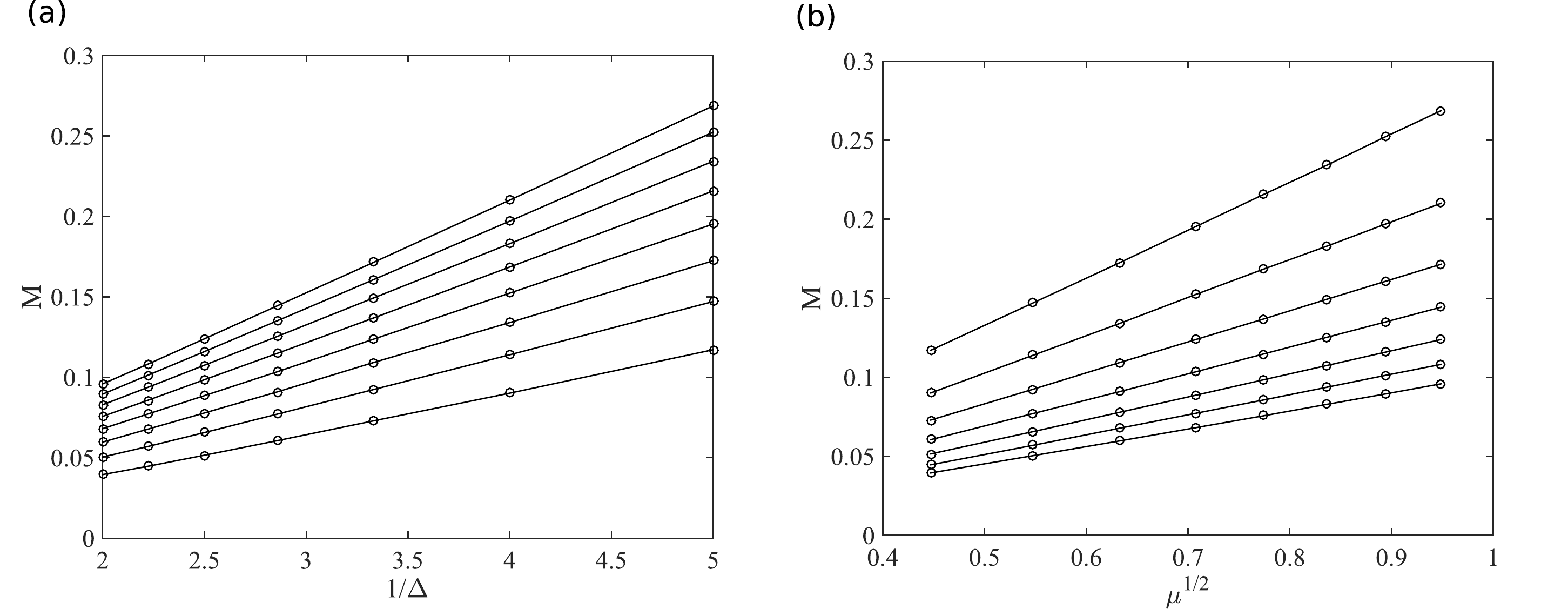}
\caption{The effective mass scales as (a) $M \propto 1/\Delta$ and (b) $M \propto \sqrt{\mu} \propto k_F$. This is allows us to relate the topological gap and the effective wall mass as $E_\text{gap} \propto k_F^2/ 2 M$. }
\label{fig:effectivemass}
\end{figure}
For a translationally invariant system with periodic boundary conditions, the excitation spectrum (plotted in Fig.~\ref{fig::BandsCriticalVelocity}) can be given as: 
\begin{equation} \epsilon'(k)_{\pm} = \pm \sqrt{ (k^2/2m - \mu )^2+ \Delta^2 k^2 } + v(t) k . \end{equation} 
From this equation, it can be seen that for $v < |\Delta|$ the spectrum is gapped (so long as $\mu \neq 0$ and $\Delta \neq 0$).
\begin{figure}[b!]
 \includegraphics[scale=0.5]{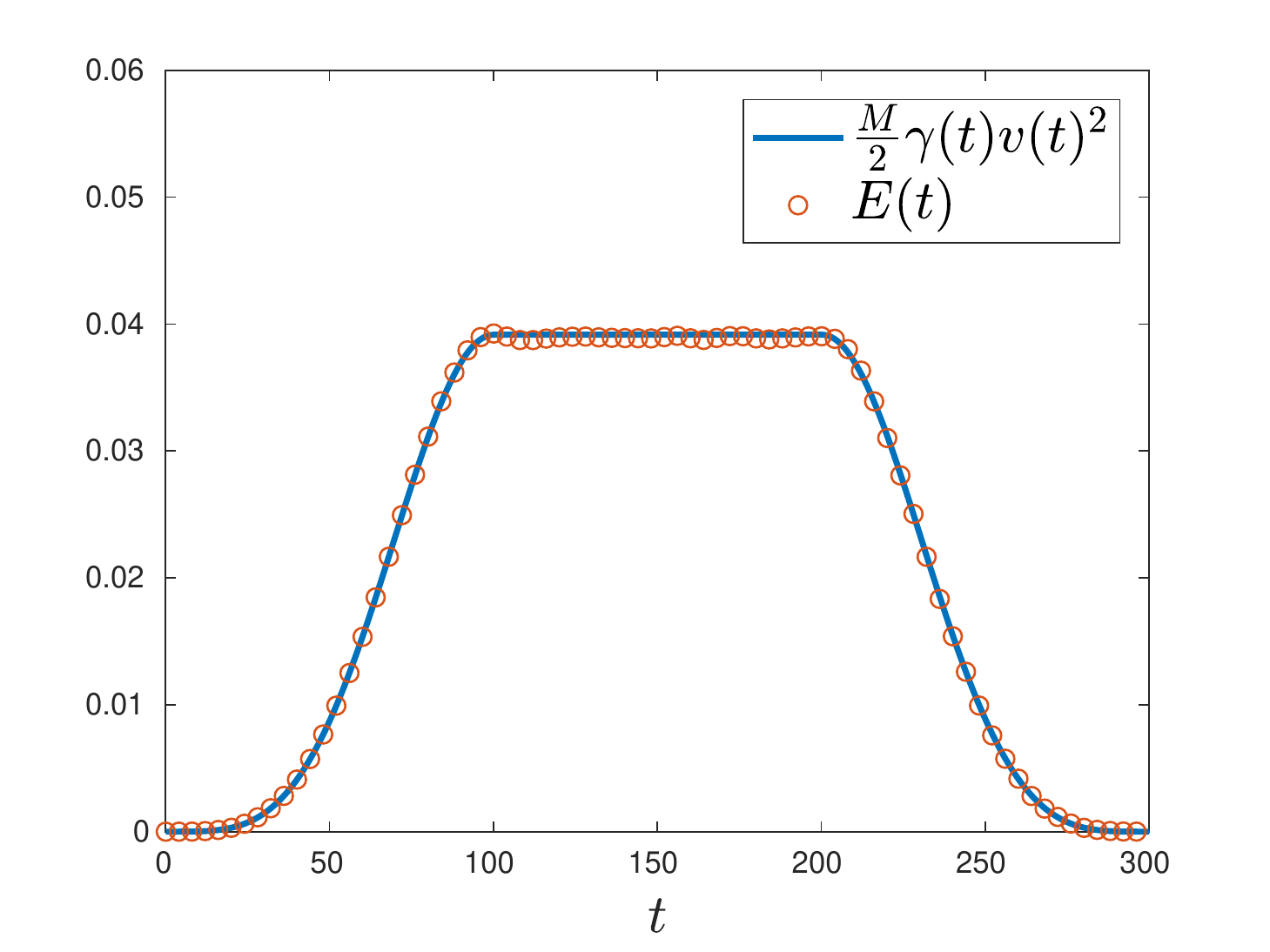}\label{fig::OverlapsEnergy}
\caption{
 (Colour online) The solid line represents the curve $\frac{M}{2}\gamma(t)v(t)^2$. The empty circles represent the value of $E(t)$ obtained through (\ref{eq:E(t)}). The value of $M\approx1.69$ was obtained by fitting the two curves. The simulations were obtained using $L=100$, $\Delta=0.4$, $\mu=1$, $a=0.5$ and $m=0.5$. The velocity profile was chosen as in (\ref{eq:v_wall_1}), with $\tau = 50$, $T = 100$ and $v_{\text{max}} = 0.2$.}
\label{fig::NumericsOverlaps}
\end{figure}
However, when $v \geq |\Delta|$ the spectrum is gapless at approximately either the positive or negative Fermi momentum $\pm k_F = \pm\sqrt{2m\mu}$. This argument suggests that there is a critical velocity at $v_\text{crit} = |\Delta|$.

Considering the system in the moving frame allows to understand intuitively why the excitations produced by the wall movement should move with momentum which is peaked approximately around $k_F$. The ground state consists of all the one particle state with $E<0$ occupied. From Fig.~\ref{fig::BandsCriticalVelocity} we see that for a sub-critical velocity $v < v_\text{crit}$, the occupied one particle state with the highest energy is at $k = -k_F$. The lowest energy unoccupied one particle state is at $k = k_F$. The smallest energy excitation between states therefore has a momentum $2k_F$. The excitations are composed by quasi-particle and quasi-hole that travel with opposite velocities (quasi-holes will eventually be reflected at the wall), which accounts for the excitations travelling with group velocity approximately peaked around $v_F$. For example, at the critical velocity $v = v_\text{crit}$ this excitation costs zero energy, and over time the topological qubit is completely lost to excitations in the wire travelling with momentum $k_F$.
\begin{figure}[t]
\includegraphics[scale=0.37]{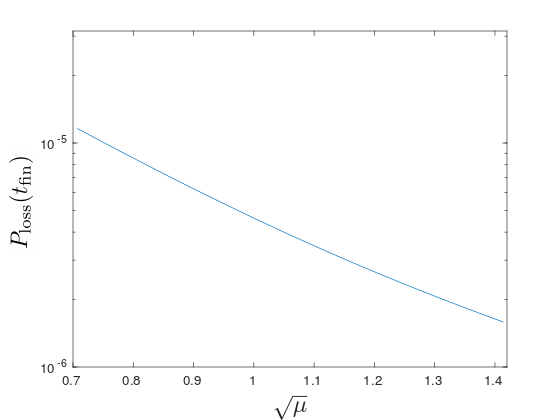}
\caption{The final qubit loss decreases exponentially in $\sqrt{\mu}$, which corresponds to moving up the electron band. This graph is produced with the lattice constant a = 0.5, the total length of the wire to be L = 100, effective electron mass m = 0.5, $\Delta$=0.2.}
\label{fig::FinalOverlapsMu}
\end{figure}

\section{Correction to the energy in the moving frame}
 \label{sect:EffectiveWallMass}

In the main text we noted that,  at the end of a movement protocol,  the accumulated deviations from super-adiabaticity results in a final resting energy above that of the ground state.
However, this result is only evident once the system has come to a complete stop.  During a super-adiabatic protocol the instantaneous energy loss can be related to a kinetic energy with an effective mass $M \propto k_F^2 / E_\text{gap} = k_{F}/\Delta$. To confirm this in the hard-wall limit we work in a position space moving frame picture and estimate the energy increase with respect to the static ground-state energy as:
\begin{equation}
\label{eq:dE_num}
 E(t) = \frac{1}{2} \text{Tr}_{+}( W_{v(t)}^\dagger H_0 W_{v(t)} - W_{0}^\dagger H_0 W_{0})
\end{equation}
\begin{figure}[t]
\includegraphics[scale=0.37]{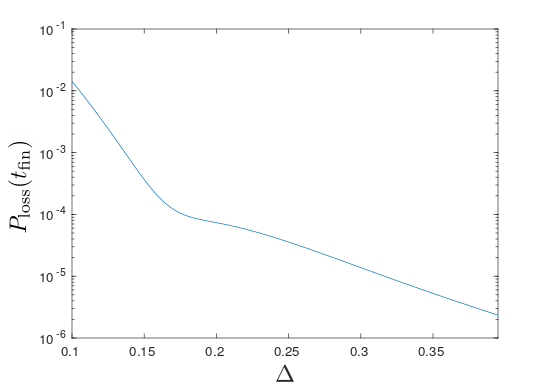}
\caption{ Analyzing final $P_{\text{loss}}$ as we vary $\Delta$ keeping $\tau = 8$ we can see past $ \Delta \sim 0.2, P_{\text{loss}}$  behaves nearly exponentially. Where $\tau$ is the length of time the system is accelerating.
This data is produced with $\mu = 1$, m = 0.5, a = 0.5 a system of length L = 100 and the lattice constant a = 0.5. } 
\label{fig::FinalOverlapsDelta}
\end{figure}
where the subscript $+$ on the trace means we only use positive energy modes $[U^T,V^T]^T$ and $W$ is defined in \ref{Wmatrix}. In general we find that if the motion is super-adiabatic then
\begin{equation}
\label{eq:E(t)}
 E(t) \approx \frac{M }{2} \gamma(t) v(t)^2.  
\end{equation}
where $\gamma(t) = 1/\sqrt{1-v(t)^2/\Delta^2}$. The effective rest mass is plotted as a function of $\Delta$ and $\mu$ in Figure \ref{fig:effectivemass}.  The scaling allows us to relate the topological gap $E_{\text{gap}} \propto k_{F}^{2}/ 2 M $. In Figure \ref{fig::NumericsOverlaps} we show a comparative plot between $E(t)$ and $\frac{M }{2} \gamma(t) v(t)^2$. The value of the effective mass is obtained by fitting methods.

\section{Qubit loss and bit flip error: additional results}
For completeness we show in Fig.~\ref{fig::FinalOverlapsDelta},\ref{fig::FinalOverlapsMu} the results of numerical analysis of the dependence of $P_{\text{loss}}(t_{\text{fin}})$ on $\Delta$ and $\mu$ for a wall movement with velocity profile as in (\ref{eq:v_wall_1}). These results are consistent with the exponential decay of the overlap  found in [\onlinecite{Scheurer2013}].

\begin{figure}[h!]
\includegraphics[scale=0.38]{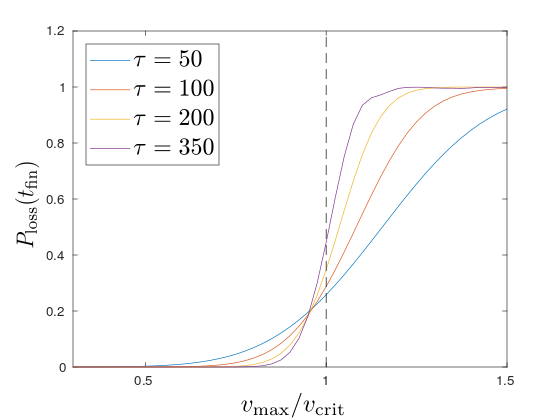}
\caption{ (Colour online) Final $P_{\text{loss}}$ for different $\tau$ and the length of the time the system continues moving at $v_{\text{max}}; T=0$. As we can see there is two different behaviours of the system depending on $v_{\text{max}}/v_{\text{crit}}$ this graph is produced with  $\Delta = 0.4$, $\mu = 1$, with a system of length $L = 100$, the lattice parameter is set to $a = 0.5$ and the effective electron mass $m = 0.5$. By examining figure (11) we can see that the different curves for $\tau$  display the same overall behaviour, in contrast to $v_{\text{max}} / v_{\text{crit}}$, where either side of $v_{\text{max}} = v_{\text{crit}}$, we can see a different effect on $P_{\text{loss}} (t_{\text{fin}})$.
 }
\label{fig::FinalOverlapsT}
\end{figure}

In Fig.~\ref{fig::FinalOverlapsT} we show $P_{\text{loss}}(t)$ at the end of the protocol for a velocity profile ($T=0$) as in (\ref{eq:v_wall_1}) and increasing acceleration times $\tau=\frac{1}{\omega}$. It can be seen that, in the limit $\tau\to\infty$, $P_{\text{loss}}(t_{\text{fin}})$ tends toward a step function in terms of $v/v_\text{crit}$. This means that for $v>v_\text{crit}$ we lose all the information stored into the ground state and it is not possible to reach the adiabatic limit. Things go differently for $v<v_\text{crit}$, where it is generally possible to reach the adiabatic limit.
In this case the instantaneous and time evolved ground state coincide at the end of the protocol, in accordance with the adiabatic theorem.

\end{document}